\documentclass{aa}  
\usepackage{xcolor}
\usepackage{txfonts,comment, amsmath}
\usepackage[colorlinks,linkcolor={blue},citecolor={blue}]{hyperref}
\usepackage{ulem}
\definecolor{blue}{rgb}{0.,0.,0.5}
\definecolor{gray}{rgb}{0.,0.5,0.}
\definecolor{darkblue}{rgb}{0,0,0.5}
\definecolor{orange}{rgb}{1.,0.5,0.}
\definecolor{darkgreen}{rgb}{0,0.5,0}

\newcommand{\Jbar}{\ensuremath{\bar J}}
\newcommand{\sline}{\ensuremath{S_{\rm L}}}
\newcommand{\kline}{\ensuremath{k_{\rm L}}}
\newcommand{\chil}{\ensuremath{\chi_{\rm L}}}
\newcommand{\chith}{\ensuremath{\chi_{\rm Th}}}
\newcommand{\ddopfid}{\ensuremath{\Delta \nu_{\rm D}^{\rm \ast}}}

\newcommand{\profilenu}{\ensuremath{\Phi_\nu}}

\newcommand\eg{\hbox{e.g.,}}

\definecolor{dylan}{rgb}{0.78, 0.08, 0.52}

\begin{document} 

   \title{Shock-heated radiation-driven outflows as a solution to the weak-wind problem of late O-type stars}
  
  \authorrunning{C. Lagae et al.}
  \titlerunning{Shock-heated radiation-driven outflows from late O-type stars}
  
   \author{C.~Lagae,
          F.~A.~Driessen,
          L.~Hennicker,
          N.~D.~Kee
          \and
           J.~O.~Sundqvist
          }

   \institute{Institute of Astronomy, KU Leuven,
              Celestijnenlaan 200D box 2401, 3001, Leuven, Belgium\\
              \email{cis.lagae@astro.su.se}
             }

   \date{Received ....; accepted ...}

 
  \abstract
   {Radiation-driven mass loss is key to our understanding of massive-star evolution. However, for low-luminosity O-type stars there are big discrepancies between theoretically predicted and empirically derived mass-loss rates (called the weak-wind problem).}   
   {We compute radiation-line-driven wind models of a typical weak-wind star to determine its temperature structure and the corresponding impact on ultra-violet (UV) line formation.}
   {We carried out hydrodynamic simulations of the line-deshadowing instability (LDI) for a weak-wind star in the Galaxy. Subsequently, we used this LDI model as input in a short-characteristics radiative transfer code to compute synthetic UV line profiles.}
   {We find that the line-driven weak wind is significantly shock heated to high temperatures and 
   is unable to cool down efficiently. This results in a complex temperature structure where more than half of the wind volume has temperatures significantly higher than the stellar effective temperature. Therefore, a substantial portion of the weak wind will be more ionised, resulting in a reduction of the UV line opacity and therefore in weaker line profiles for a given mass-loss rate. Quantifying this, we find that weak-wind mass-loss rates derived from unsaturated UV lines could be underestimated by a factor of between 10 and 100 if the high-temperature gas is not properly taken into account in the spectroscopic analysis. This offers a tentative basic explanation for the weak-wind problem: line-driven weak winds are not really weaker than theoretically expected, but rather a large portion of their wind volume is much hotter than the stellar effective temperature.}
   {}


   \maketitle
%

\section{Introduction}
The lives and deaths of massive stars are strongly influenced by the huge amount of mass loss they experience during their evolution in the form of radiation-driven stellar winds \citep{Smith14}. During (at least) the massive-star main sequence, the driving force of the stellar winds originates from resonance scattering by stellar radiation in spectral lines. The first quantitative description of such line driving was made by \cite{CAK} (hereafter CAK), in which the wind was assumed to be smooth and steady. Extensions to this CAK theory \citep{FA86,Pauldrach86} have been relatively successful in explaining the basic properties of these line-driven winds, such as the observed metallicity dependence of mass-loss rate \citep{mokiem07b} and the relation between stellar luminosity and wind momentum \citep{Kud95}.

Further theoretical investigations of line-driven winds by \citet{MacGregor79}, \citet{Carlberg80}, and \cite{ORI} concluded, in contrast to CAK theory, that these winds are subject to a small-scale radiation instability inherent to the line driving, the so-called line-deshadowing instability (LDI). One-dimensional numerical radiation-hydrodynamic simulations \citep[e.g.,][]{OCR88, Feldmeier1995, driessen19} following the non-linear evolution of the LDI have shown that these perturbations grow into small dense shells separated by a nearly void, rarefied medium. Following this, 2D simulations  \citep{DO03,Sundqvist2018} showed that these shells break up laterally into dense clumps with a size that is on the order of the Sobolev length 
(typically only 1\%\ of the stellar radius), which is the length scale over which the wind accelerates one thermal speed \citep{Sobo60}.

Accounting properly for such wind clumping is essential for correctly estimating mass-loss rates from observational spectral diagnostics such as ultra-violet (UV) resonance and H$\alpha$ lines \citep{Puls06,Sundqvist2018}. When wind clumping is properly included in the analysis, empirically derived mass-loss rates typically agree well with the newest theoretical predictions for early O-stars with relatively dense winds \citep{Bjorklund20}. On the other hand, observations of Galactic late-O main sequence stars with lower-density winds \citep{martins04,martins05b, Marcolino09, Almeida19} have revealed a trend where empirically derived mass-loss rates are consistently much lower than predictions from theoretical models. Specifically, for low-luminosity stars in the Galaxy $(L_\star < 10^{5.2} L_\sun)$ the empirically derived mass-loss rates can be up to two orders of magnitude lower than predicted by the models from \citet{Vink01}, which means they are also substantially lower than the newer predictions by \citet{Bjorklund20}. \cite{bouret03} defined these weak-wind stars as having mass-loss rates of $\dot{M}<10^{-8} M_\sun/\mathrm{yr}$, thus residing in a regime where the optical H$\alpha$ line is insensitive to mass loss and the strong UV wind lines, the so-called P-Cygni lines, thereby become the primary diagnostics for inferring empirical mass-loss rates (though see \citealt{Najarro11} for a potential infrared (IR) alternative). Besides the discrepancy in mass-loss rates, lower terminal wind speeds have also been derived from observations compared to theoretical models \citep{Sundqvist19}. These aforementioned differences between theory and observations are typically summarised as the weak-wind problem.
This weak-wind problem is most often discussed in the context of O-type stars, leaving the domain of even less luminous B-type stars for the most part unexplored. While empirical mass-loss rate calculations \citep{Cohen97, Ryspaeva19} and theoretical mass-loss rate predictions \citep{Krticka14} have been carried out independently for B-stars, to our knowledge no detailed study similar to those performed for O-stars has been made  so far. Whether or not early B-type stars are subject to the weak-wind problem in the same sense as late O-stars therefore remains an open question.
 
The mass-loss rate  discrepancies observed in the weak-wind regime strongly suggest that for such low-luminosity O-type stars there is underlying physics that is altering either the observational wind diagnostics used to derive mass-loss rates, the theoretical description of line-driven winds, or both (see overview by \citealt{Puls08b}). Of particular importance for our work here, \cite{Lucy2012} proposed a phenomenological wind model wherein weak winds are shock heated to coronal temperatures ($T>10^6$ K). Because of the low density of the weak winds, they would subsequently be unable to cool down the shock-heated zones such that these remain hot and advect outwards. This general picture would be in relatively stark contrast to the standard model for embedded shocks in line-driven winds \citep{Feldmeier95, Feldmeier97a}. Namely, within this standard scenario the strong shocks caused by the LDI cool down very efficiently by radiation, leaving only a very small percentage of the total wind volume (typically a few percent) hotter than the stellar effective temperature. However, as all LDI models thus far have been computed for stars with mass-loss rates well above $10^{-8} M_\sun/\mathrm{yr}$, it remains to be investigated how LDI-generated shocks develop in the weak-wind regime. Indeed, although previous LDI models have been successful in explaining X-ray emission from dense O-supergiants \citep{Feldmeier97a}, X-ray observations of late O-type stars \citep{Huenemoerder12,Doyle17} and early B-type stars \citep{Cazorla17} do seem to indicate that these winds might have larger portions of hot gas embedded in them. 

Therefore, here we undertake an investigation of how shock heating from the LDI impacts the global temperature structure of weak winds, and attempt to decipher whether or not this might offer a basic explanation for the weak-wind problem. Building on the hypothesis by Lucy, the basic idea is that the low densities of weak winds make cooling very inefficient. This may be a very reasonable argument because radiative cooling scales with density squared, and cooling by adiabatic expansion is typically negligible in these supersonic line-driven flows. As the LDI then shock heats the wind to coronal temperatures, it will become increasingly ionised and therefore might consist of large hot regions in which the opacities of specific UV line transitions (e.g. those of C\,IV) are significantly reduced. This would mean that synthetic predictions of such UV resonance lines would be reduced accordingly, thus also affecting the empirical mass-loss determinations discussed above. 

We compute wind models of a typical weak-wind star to investigate the aforementioned hypothesis and compare its wind structure to a well-understood O-supergiant model. In addition, a wind model for a late B-type dwarf is computed to also explore this very low-density wind regime. In $\S$\ref{Sec:method} we develop numerical radiation-hydrodynamic simulations of the non-linear evolution of the LDI in low-density winds of late O-type and early B-type stars. 
We carefully analyse the resulting structure and how it affects wind properties and line profiles in $\S$\ref{Sec:results}. Lastly, in $\S$\ref{Sec:conclusion} we discuss our findings, summarise key results, and detail future work.

\section{Method}\label{Sec:method}
\subsection{Radiation hydrodynamics} 
The theoretical wind models in this work are computed using the parallel adaptive mesh refinement code MPI-AMRVAC \citep{Xia2018}. We use this flexible code framework to solve the fluid conservation equations on a 1D spherical grid with additional user-written modules from \cite{driessen19} for computation of the LDI line force (see their Appendix A  for details). Specifically, we solve the 1D spherically symmetric hydrodynamic equations including radiation effects using a HLLC solver \citep{Toro94} together with a Piecewise Parabolic slope limiter \citep[PPM]{Colella1984}:
\begin{equation}
\begin{split}
   &\frac{\partial\rho}{\partial t} + \frac{1}{r^2}\frac{\partial(r^2\rho \varv)}{\partial r} = 0~~,\\
   &\frac{\partial\rho \varv}{\partial t} + \frac{1}{r^2}\frac{\partial(r^2\rho \varv\varv)}{\partial r} = -\frac{\partial P}{\partial r} + \rho g_\mathrm{line} + \rho g_\mathrm{eff}~~,\\   
   &\frac{\partial e}{\partial t} + \frac{1}{r^2}\frac{\partial(r^2 e\varv)}{\partial r} = -\frac{P}{r^2}\frac{\partial(r^2 \varv)}{\partial r} - C_\mathrm{rad}~~,\\      
\end{split}
\end{equation}
where $\rho$, $\varv$, and $e$ are the density, velocity, and total energy density, respectively. Here $g_\mathrm{line}$ is the line force due to resonance line scattering (further detailed in $\S$\ref{Subsec line-force}) and:
\begin{equation}\label{eq: geff}
    g_\mathrm{eff} = -\frac{GM_\star(1-\Gamma_\mathrm{e})}{r^2}
\end{equation}
is the effective gravity accounting for electron scattering with Eddington factor $\Gamma_\mathrm{e} \equiv \kappa_e L_\star /(4\pi GM_\star c)$ and electron-scattering opacity $\kappa_e = 0.34$ cm$^2$ g$^{-1}$.
In the energy conservation equation, the two terms on the right describe energy losses due to adiabatic expansion and radiative cooling, respectively. The total energy density is related to the gas pressure as the sum of internal and kinetic energy:
\begin{equation}\label{eq: gline}
    e = \frac{P}{\gamma-1} + \frac{\rho \varv^2}{2},
\end{equation}
with adiabatic index $\gamma = 5/3$.

\subsection{Line force}\label{Subsec line-force}
The main challenge when computing our wind models is the calculation of the line force in the absence of the Sobolev approximation \citep{Sobo60}. We follow the escape probability method of \cite{OP96} to compute the 1D line force for an ensemble of lines. Here we only give a brief overview of the line-force parameters and ensemble; a full description is given in Appendix A of \cite{driessen19}. The radiative line
force is defined as the sum of a direct and diffuse component given by equations $(\mathrm{A.}4)$ and $(\mathrm{A.}5)$ in \citeauthor{driessen19}, respectively. The line ensemble enters the line
force through eq. ($\mathrm{A.}3$) in \citeauthor{driessen19}. It is modelled as a power-law distribution wherein the line distribution is exponentially truncated at a maximum line strength $Q_\mathrm{max}$:
\begin{equation}
    \frac{\mathrm{d}N}{\mathrm{d}q} = \frac{1}{\Gamma(\alpha)\Bar{Q}}\left( \frac{q}{\Bar{Q}} \right)^{\alpha - 2} e^{-q/Q_\mathrm{max}}~~,
\end{equation}
where $\alpha$ is the CAK power-law index, $q$ the line strength according to the formalism developed by \cite{Gayley95}, and $\Bar{Q}$ is a line-force normalisation constant that describes the ratio between the total line
force and the force due to electron scattering in the case where all lines are optically thin. 

Typical values of O-supergiants at solar metallicity are $\Bar{Q} \approx Q_\mathrm{max}\approx2000$ \citep{Gayley95, Puls00}. However, due to numerical limitations, the line strength cut-off is often decreased to $Q_\mathrm{max}\approx 0.004\Bar{Q}$ \citep{OCR88}. In the case of low-density winds from late O- and early B-type stars we are able to increase $Q_\mathrm{max}$ to more realistic values $Q_\mathrm{max}\approx 0.4\Bar{Q}$ due to the lower overall opacity in the wind. Table \ref{tab:parameters} summarises the stellar and wind parameters used in this paper for three representative stars in the Galaxy. The stellar parameters for the weak wind O-dwarf are based primarily on HD\,326329 from Table 3 in \citet{Marcolino09} while the B-dwarf is based on $\delta$\,Cen from Table 2 in \citet{Cohen97}. Lastly, the temporal averaged mass-loss rates from our LDI simulations are also shown in Table \ref{tab:parameters}. It is important to point out here that the goal of the LDI simulations is to analyse the resulting wind structure, and not to predict mass-loss rates. As such, the specific values for the line-force parameters have simply been chosen such that we obtain average mass-loss rates that are similar to theoretical predictions based on full line-list calculations \citep[e.g.][]{Vink01,Bjorklund20}.

\begin{table}[]
    \centering
    \caption{Overview of adopted stellar and wind parameters.}
        \def\arraystretch{1.5}
    \begin{tabular}{llll}
    \hline \hline
        Variable & O-supergiant & O-dwarf & B-dwarf  \\ \hline
        $M$ [$M_\sun$]& $50$ & $20$ & $10$ \\
        $T_\mathrm{eff}$ [K] & $40\,000$ & $30\,000$ & $23\,000$ \\
        $L$ [$L_\sun$] & $8\times10^5$ & $3.6\times10^4$ & $6\,310$ \\
        $R$ [$R_\sun$] & 20 & 7 & 5 \\
        $\langle\dot{M}\rangle$ [$\dot{M}_\sun\mathrm{yr}^{-1}$] & $2.3\times10^{-6}$ & $2.4\times10^{-8}$ & $1.7\times10^{-9}$ \\
        $\alpha$ & 0.65 & 0.65 & 0.65 \\
        $\Bar{Q}$ & 2000 & 2000 & 2000 \\ \hline
    \end{tabular}
    \label{tab:parameters}
\end{table}

\subsection{Radiative cooling}
Optically thin radiative cooling is implemented following the exact integration scheme from \cite{Townsend09} together with the SPEX cooling curve from \cite{Schure09}. The energy loss due to radiative cooling enters the energy conservation equation as a term proportional to density squared:
\begin{equation}
    C_\mathrm{rad} = -\rho^2\Lambda_\mathrm{m}(T),
\end{equation}
where $\Lambda_\mathrm{m}$ is the mass-weighted energy loss per second at temperature $T$ (called the cooling curve). Additionally, a floor temperature is set equal to the stellar effective temperature to mimic photo-ionisation heating from the stellar radiation field \citep[see also][]{Runacres02, Sundqvist13}.

To test the addition of radiative cooling in our models, we simulated the extreme case of a low-density wind from a B-dwarf. From these simulations we find that numerical cooling in the inner wind regions ($r<2R_\star$) seems to be much more efficient than theoretically expected. Following the individual (temperature) evolution of clumps is difficult in an Eulerian framework. However, we confirm our hypothesis by calculating the time needed to cool down to floor temperature, namely the cooling time:
\begin{equation}
   t_\mathrm{c} = -\frac{3k}{2\Bar{\mu}\rho}\int_T^{T_\mathrm{floor}}\frac{\mathrm{d}T}{\Lambda_\mathrm{m}(T)}~~,
\end{equation}
with $k$ the Boltzmann constant and $\Bar{\mu}$ the mean atomic mass, in every cell of our grid. Cooling is expected to be efficient if the local cooling time is shorter than the local dynamical timescale, which is approximated here as:  
\begin{equation}
    t_\mathrm{dyn} = R_\star/\varv(r)~~.
\end{equation}
In the opposite case, $t_\mathrm{c}>t_\mathrm{dyn}$, the parcel of gas is expected to stay hot and should not cool down significantly during its outwards evolution. Furthermore, because the average wind velocity monotonically increases while the average density monotonically decreases with distance to the star, the fraction $t_\mathrm{c}/t_\mathrm{dyn}$ is a lower limit.

\begin{figure}[ht]
  \resizebox{\hsize}{!}{\includegraphics{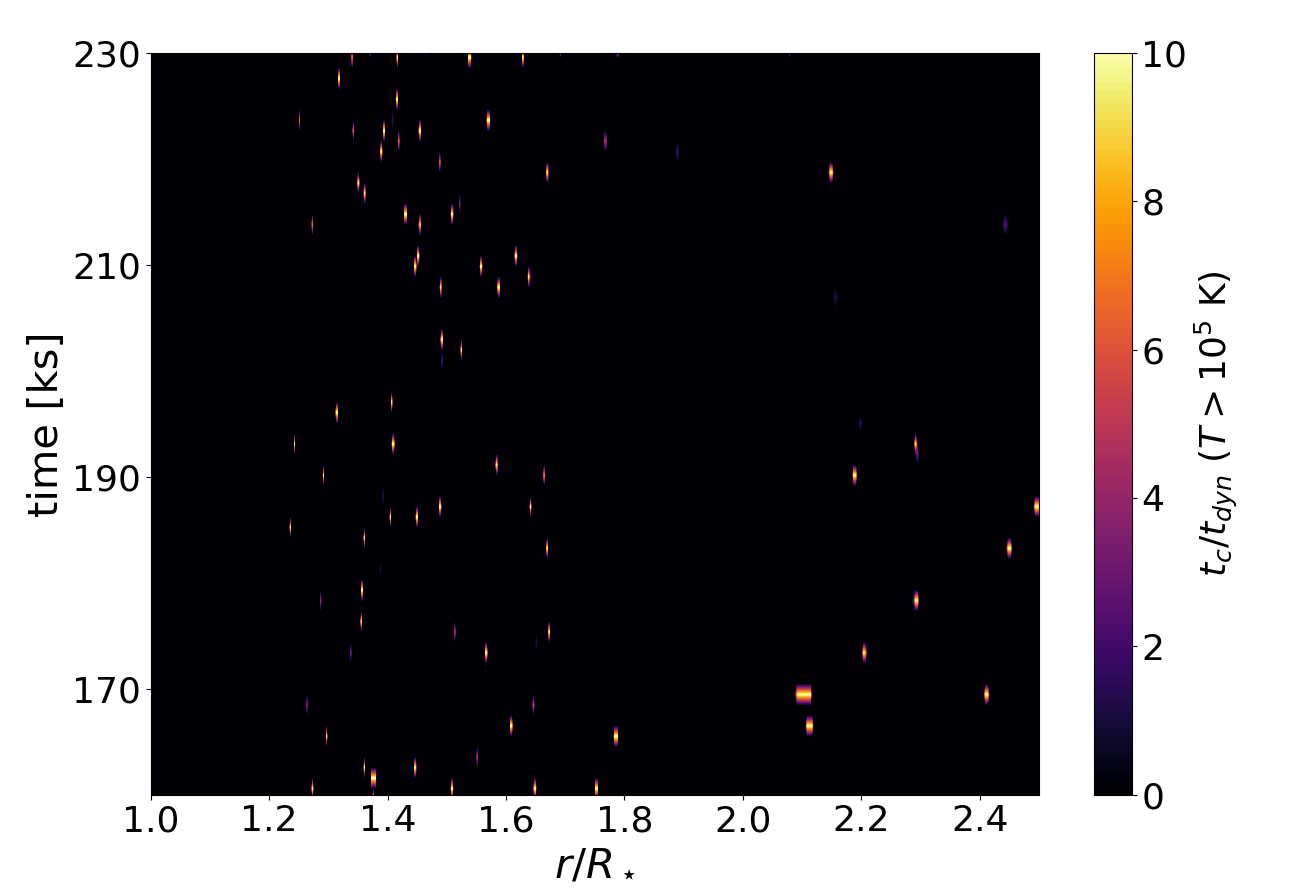}}

\caption[]{Colour plot of the temporal evolution of the ratio $t_\mathrm{c}/t_\mathrm{dyn}$ in the inner wind, where only gas parcels hotter than $10^{5}$ K are shown. For visualisation purposes, the maximum value of $t_\mathrm{c}/t_\mathrm{dyn}$ is set to ten.}
\label{fig:tc_frac}
\end{figure}

Figure \ref{fig:tc_frac} shows a spatio-temporal temperature map from a subset of our test simulation, revealing the presence of many hot gas parcels at $r\approx 1.5R_\star$ that should theoretically not cool down because $t_\mathrm{c}/t_\mathrm{dyn} = 2-10$ there. Nevertheless, barely any hot gas remains one snapshot later at slightly larger radii $r \approx 1.7-2.2R_\star$, where we would expect the hot parcels from the previous snapshot to be visible. Here, we attribute this absence of hot gas parcels to inherent problems with numerical overcooling. Such numerical overcooling issues are well known to occur in simulations of the LDI, and have been discussed ever since the first non-isothermal model by \cite{Cooper92}. Moreover, this problem affects not only simulations of the LDI, but is quite a general issue of supersonic flow simulations where radiative cooling is important. For example, \cite{Parkin10} and \cite{Steinberg18} demonstrate that numerical overcooling will in principle always occur in Eulerian mesh codes because of advective diffusion at shock fronts. To circumvent this, the non-isothermal LDI simulations by \cite{Feldmeier95} applied a correction to the cooling curve, forcing it to follow a power law, $\Lambda(T) \propto T^2$ , below a certain cut-off temperature, thereby alleviating the above-discussed problems. \citet{Feldmeier97a} then applied this model to simulate the LDI for an O-supergiant wind, showing that the amount of hot gas in the wind then indeed becomes comparable with observations. In the spirit of performing a comparative study of LDI-generated structures in such dense O-supergiant winds and in more low-density OB-type star winds, all simulations presented in $\S$\ref{Sec:results} of this paper apply the same correction to the cooling curve as \citet{Feldmeier95}. This then allows us to perform a differential study, investigating properties of low-density winds while simultaneously following the well-established standard method for simulating denser O-supergiant outflows. 

\subsection{Initial and boundary conditions}\label{subsec: boundary conditions}
All models are initialised from a CAK-type wind simulated on a non-uniform grid spanning ten stellar radii ($1 - 10R_\star$). The CAK wind itself is initialised from a $\beta$-velocity law and allowed to relax by evolving the wind until it settles into a smooth and steady state. The numerical grid contains 2000 cells that are stretched according $\Delta r_\mathrm{i+1} = 1.002\Delta r_\mathrm{i}$ in order to obtain a higher resolution close to the subsonic base. The latter is necessary to resolve the steep acceleration region surrounding the sonic point. At the inner boundary, we fix the density to roughly four to five times the density at the sonic point \citep{Sundqvist13}, and set the velocity by a constant extrapolation of the slope. Additionally, the boundary density of the CAK model is tuned to obtain a hydrodynamic mass loss rate, $\dot{M} = 4\pi\rho\varv r^2$, that lies within a factor of two of the analytical CAK mass-loss rate while still allowing the wind to relax. At the supersonic outer boundary, we take no-inflow conditions (extrapolation assuming constant gradients) for the hydrodynamic variables such that no spurious wave feedback happens from the outer boundary to the mesh.

Each simulation is evolved for at least $25t_\mathrm{dyn}(\varv_\infty)$ corresponding to a total simulation time of $t>300~\mathrm{ks}$. To ensure numerical stability, the time-step is chosen as the minimum between a Courant–Friedrichs–Lewy time-step \citep[CFL]{CFL28} and a user-defined time-step:
\begin{equation}
    \Delta t = \mathrm{min}\left( \Delta t_\mathrm{CFL}, 0.3\sqrt{\frac{\Delta r}{g_\mathrm{line}+g_\mathrm{eff}}} \right)~~,
\end{equation}
with $g_\mathrm{line}$ and $g_\mathrm{eff}$ the line
force (see \citealt{driessen19}) and effective gravity (eq. \ref{eq: geff}) , respectively.

\subsection{Post-processed line-profile computations} 
\label{subsec:profiles}
An important application of the low-density wind models developed in this paper is to investigate how the LDI-generated structure affects the formation of UV resonance lines. As discussed in Sect. 1, such P-Cygni lines are typically used to derive observational constraints on wind terminal velocities and mass-loss rates of OB-type stars. 

To compute synthetic UV line profiles from our simulations, we use the 3D short-characteristics radiative transfer code developed by \cite{Hennicker2020}. This code is formulated in Cartesian coordinates and applies a `star-in-a-box' model to calculate resonance-line transitions formed in the stellar wind. The time-independent equation of radiative transfer is solved in the frame of the  observer, enabling us to account for non-monotonic (albeit non-relativistic) velocity fields. The solution scheme for the discretised equation of radiative transfer (see \citealt{Hennicker2020}, their eq. 12) then requires boundary conditions, specified here by setting 
$I_\nu^+=B_\nu(T_{\mathrm{eff}})$ and $I_\nu^-=0$, where $B_\nu(T_\mathrm{eff})$ is the Planck function evaluated at the stellar effective temperature, and $I_\nu^+$, $I_\nu^-$ describe the intensities emerging from the stellar core and entering the wind from outside, respectively. The line is approximated by a two-level atom, with the opacity parameterised by
\begin{equation}
   \chil=\kline \chith \ddopfid \profilenu \,,
\end{equation}
where $\kline$ is the line-strength parameter (see, ~\citealt{Hennicker2018}), $\chith\propto\rho$ is the Thomson opacity (in $[1/\mathrm{cm}]$) depending on density (and thus also on the mass-loss rate $\dot{M}$ via the equation of continuity), $\ddopfid$ is a fiducial Doppler width (not further specified here), and $\profilenu$ is the profile function of the line transition. As the line opacity depends linearly on density, the absorption part of a UV resonance line transition can be used to infer mass-loss rates. Further, $\kline$ is used as an input parameter within our calculations (rather than calculated from the quantum mechanical properties of the transition). Thus, an increase or decrease in $\kline$ is directly related to an increase or decrease in the mass-loss rate, respectively.

Throughout this paper, we consider a pure scattering line for which the line source function depends on the radiation field and can be written as $\sline=\Jbar$, with $\Jbar$ the profile-weighted, frequency-integrated, and angle-averaged specific intensity (thus depending on the source function via the equation of radiative transfer). To solve this coupled problem, the 3D code applies an accelerated $\Lambda$-iteration scheme based on \citealt{Cannon73} and \citealt{Scharmer1981} (see also \citealt{Hennicker2020}, Sect. 3.6, for the corresponding formulation within the 3D short-characteristics method applied here). The emergent flux profile is then calculated using the converged source function within a post-processing long-characteristics solution method (see, \eg~\citealt{Lamersetal87}, \citealt{Busche2005}, \citealt{Sundqvist12c}, and \citealt{Hennicker2020}). To this end, a cylindrical coordinate system is defined with the $z$-axis being aligned with the line-of-sight towards the observer. While the intensities are calculated by solving the equation of radiative transfer along rays in the $z$-direction, the emergent flux is obtained by integrating the emergent intensities over the projected stellar disc (including the wind region).
 
To investigate how the shock-heated gas affects the line formation, we simply set the opacity to zero in all regions with temperatures $T > 10^5$ K, because at higher temperatures the elements of typical key diagnostic lines in the UV such as C\,III/C\,IV, N\,IV/N\,V, Si\,IV, and P\,V are expected to reside in significantly higher ionisation stages. 
Moreover, to allow for a full 3D description, we select a number of $N$ random 1D snapshots from the corresponding LDI simulation and `patch' these together in latitude and azimuth. Although this method only produces a `pseudo-3D' wind structure, it serves as a first approximation of the lateral break-up of high-density shells expected in truly multi-dimensional radiation-hydrodynamic simulations of the LDI \citep{Sundqvist2018}. All profiles shown in $\S$\ref{Subsec:line profiles results} are computed assuming $N = 5151$; however, by increasing and decreasing $N$ to 20301 and 1225 we verified that this has a marginal impact on the resulting line profiles. 

\section{Results}\label{Sec:results}
\subsection{Hydrodynamic wind models}
\begin{figure*}[ht]
  \resizebox{\hsize}{!}{\includegraphics{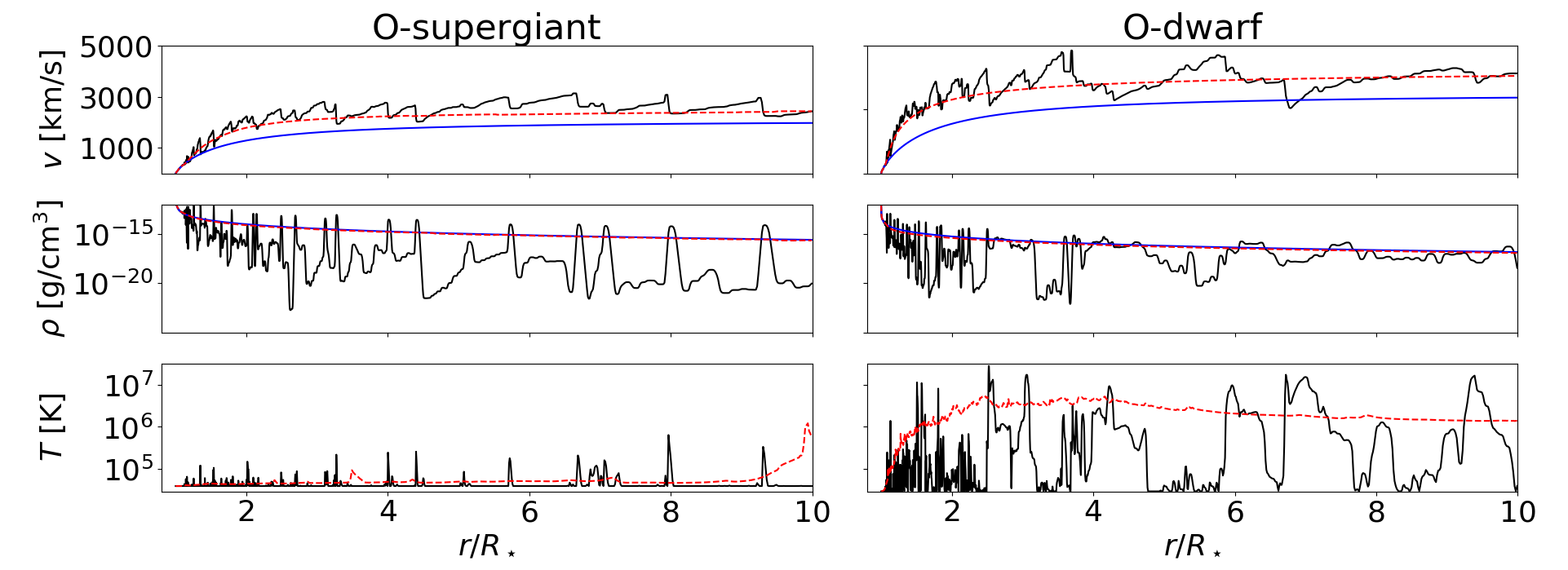}}

\caption[]{Snapshot of the radial velocity, density, and temperature profiles for the O-supergiant (left) and O-dwarf (right) wind models. The blue lines indicate the corresponding initial CAK models and the red dashed lines are temporal averages taken at each grid cell.} 
\label{fig:RadialProf}
\end{figure*}
\begin{figure}[ht]
  \resizebox{\hsize}{!}{\includegraphics{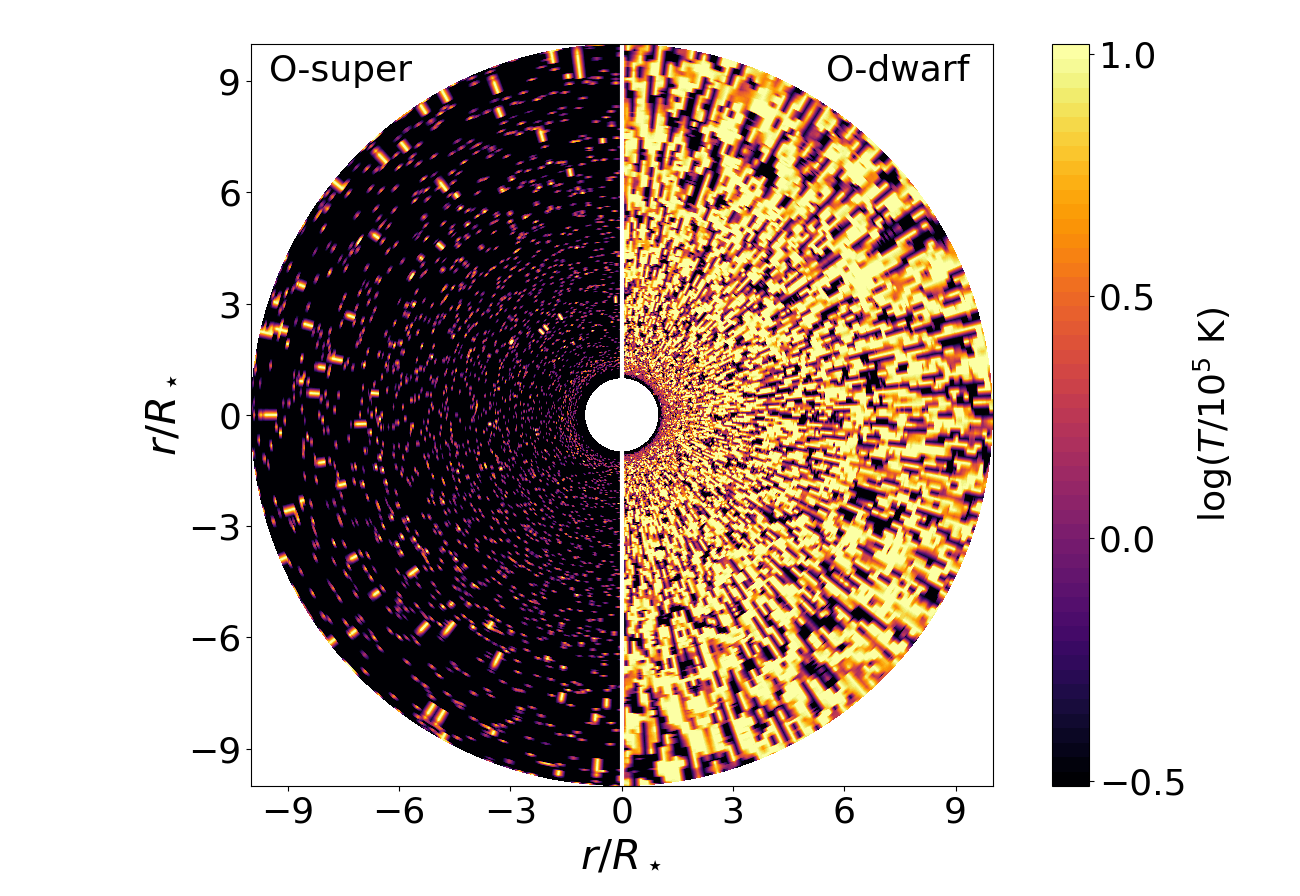}}

\caption[]{Temperature contour plot for the O-supergiant (left half) and O-dwarf (right half) wind model. The temperature is normalised to $T=10^5$ K and artificially limited to a maximum value of $10^6$ K for visualisation purposes.}
\label{fig:OBtemp}
\end{figure}
\begin{figure}[ht]
  \resizebox{\hsize}{!}{\includegraphics{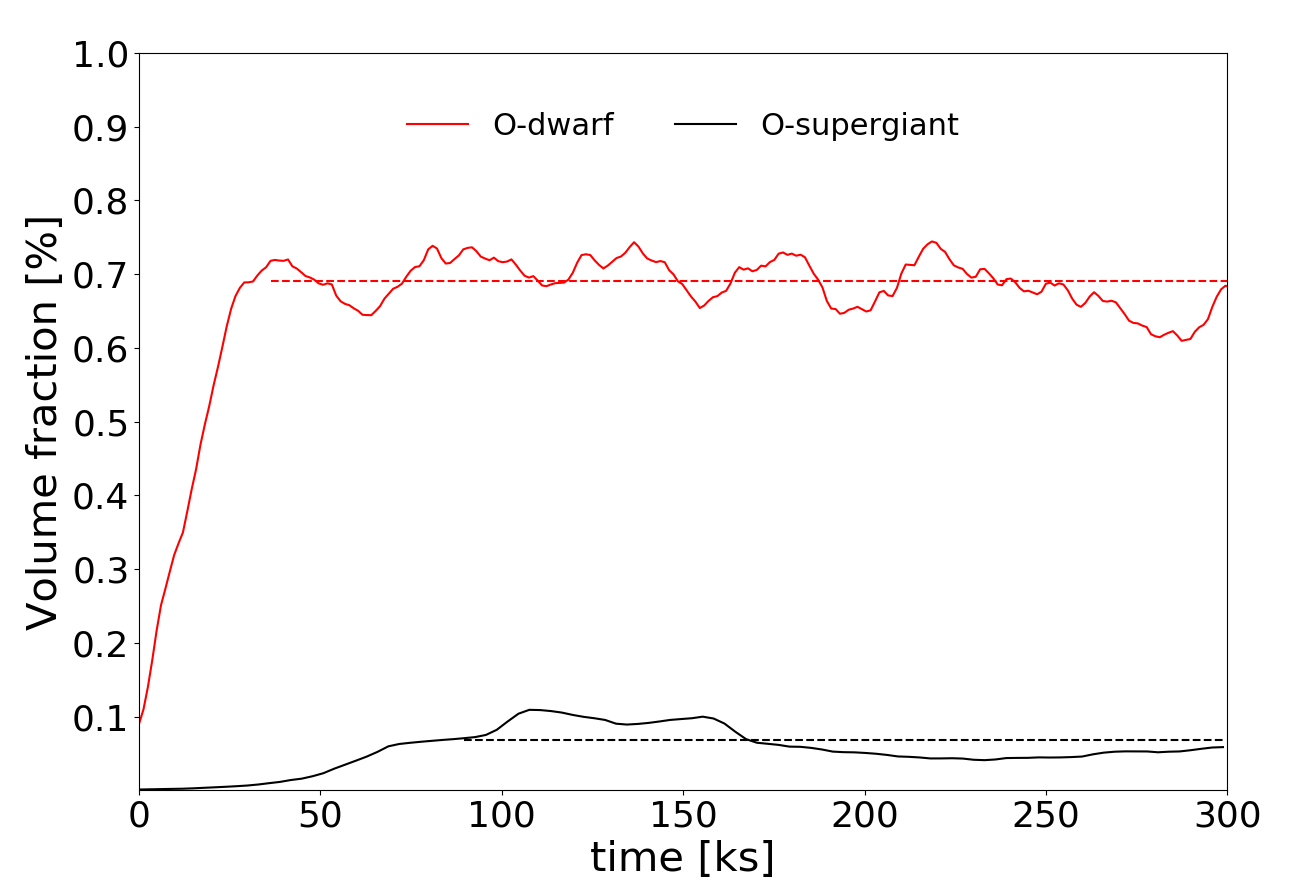}}

\caption[]{Temporal evolution of the volume fraction of hot gas ($T>10^5$ K) in the O-supergiant (black) and O-dwarf (red) wind model. The average value over time is shown as a dashed line for each model.}
\label{fig:Vfrac}
\end{figure}

A snapshot of the radial velocity and density profile for the low-density O-dwarf and reference O-supergiant wind models, together with their averages and initial smooth CAK condition, are shown in Fig. \ref{fig:RadialProf}. The profiles show clear differences between the two models. First, clumps are broader and less pronounced in the low-density wind of the O-dwarf, and the interclump medium is significantly less rarefied at larger distances from the stellar surface compared to the reference O-supergiant model. 
Secondly, and perhaps of greater importance for our work here, the O-dwarf wind contains multiple broad regions with high temperatures. Already quite close to the stellar surface, the average temperature rises above $10^5$ K after which it reaches values of $10^6$ K throughout the rest of the wind.
To obtain a visual overview of the volume occupation of these hot regions, a 2D contour plot is shown in Fig. \ref{fig:OBtemp} for both the O-dwarf and reference O-supergiant model. This figure is composed of individual random snapshots of the 1D simulation placed at varying angles to fill the circle, and visually reveals that a substantial portion of the O-dwarf wind resides at temperatures significantly above the stellar effective temperature. 
Further quantitative analysis of the volume fraction of hot gas (i.e. at $T>10^5$ K) in the wind is shown in Fig. \ref{fig:Vfrac}. These percentages are calculated by summing over the volume of all cells that contain hot gas and dividing it by the total wind volume. Doing so, we find that on average $69\pm10\%$ of the O-dwarf wind is hot, whereas this value is only $7\pm5\%$ in the template O-supergiant simulation.

In addition to the O-dwarf, we also modelled the very low-density wind from a B-dwarf (see Table \ref{tab:parameters}), for which we find that the volume fraction of hot gas is on average $39\pm8\%$. Compared to the O-dwarf and -supergiant wind, this value lies roughly in the middle. We attribute the higher fraction of hot gas for the O-dwarf to a combination of its low wind density and the presence of strong shocks. In particular, although the wind density is even lower for the B-dwarf, the shocks  there are significantly weaker. Overall, it appears that the O-dwarf indeed lies in a very good parameter range for obtaining a large volume fraction of hot gas, because this requires a combination of sufficient heating (strong shocks) and inefficient cooling (low wind density).

The obtained results have two important physical consequences for low-density line-driven winds, because these hot regions will be mostly transparent to stellar radiation such that light can more easily leak through. Firstly, the reduction in line opacity already near the stellar surface may decrease the line-driving force. The effect of this reduction is only observed when comparing an isothermal model to the model shown in Fig. \ref{fig:RadialProf}, which gives only a small but noticeable reduction in terminal wind speed. However, we note that this reduction in line opacity for hot gas has been modelled  here by applying a very simple cut-off in the LDI line opacity as a function of temperature, whereas the line driving actually depends on the detailed wind ionisation balance and the presence of various line-driving species. However, it is currently not possible to include such a detailed calculation in a time-dependent LDI simulation.

Secondly, and more important for our purposes here, the hot wind regions will substantially affect critical spectral diagnostics such as UV resonance lines. Namely, because a substantial portion of the wind will now reside in significantly higher ionisation stages, the opacities of key strategic lines for inferring empirical mass-loss rates will be significantly reduced as compared to what is assumed in stellar atmosphere codes used for quantitative spectroscopy. Accordingly, the line opacities of higher ionisation stages can be expected to increase, with number densities presumably distributed over several ionisation stages because of the wide range of temperatures within the shock-heated material. We therefore speculate that the corresponding lines might only barely be visible. 
However, investigation of this effect in further detail would require the determination of the ionisation balance accounting for a large number of elements and line transitions within the 3D clumped wind. As such calculations are currently not feasible in terms of computation time (if the corresponding methods existed at all), we focus on the aforementioned effect of reduced opacities for a typical ionisation stage in the following.

\subsection{Line profiles}\label{Subsec:line profiles results}
\begin{figure}[ht]
%
  \resizebox{\hsize}{!}{\includegraphics{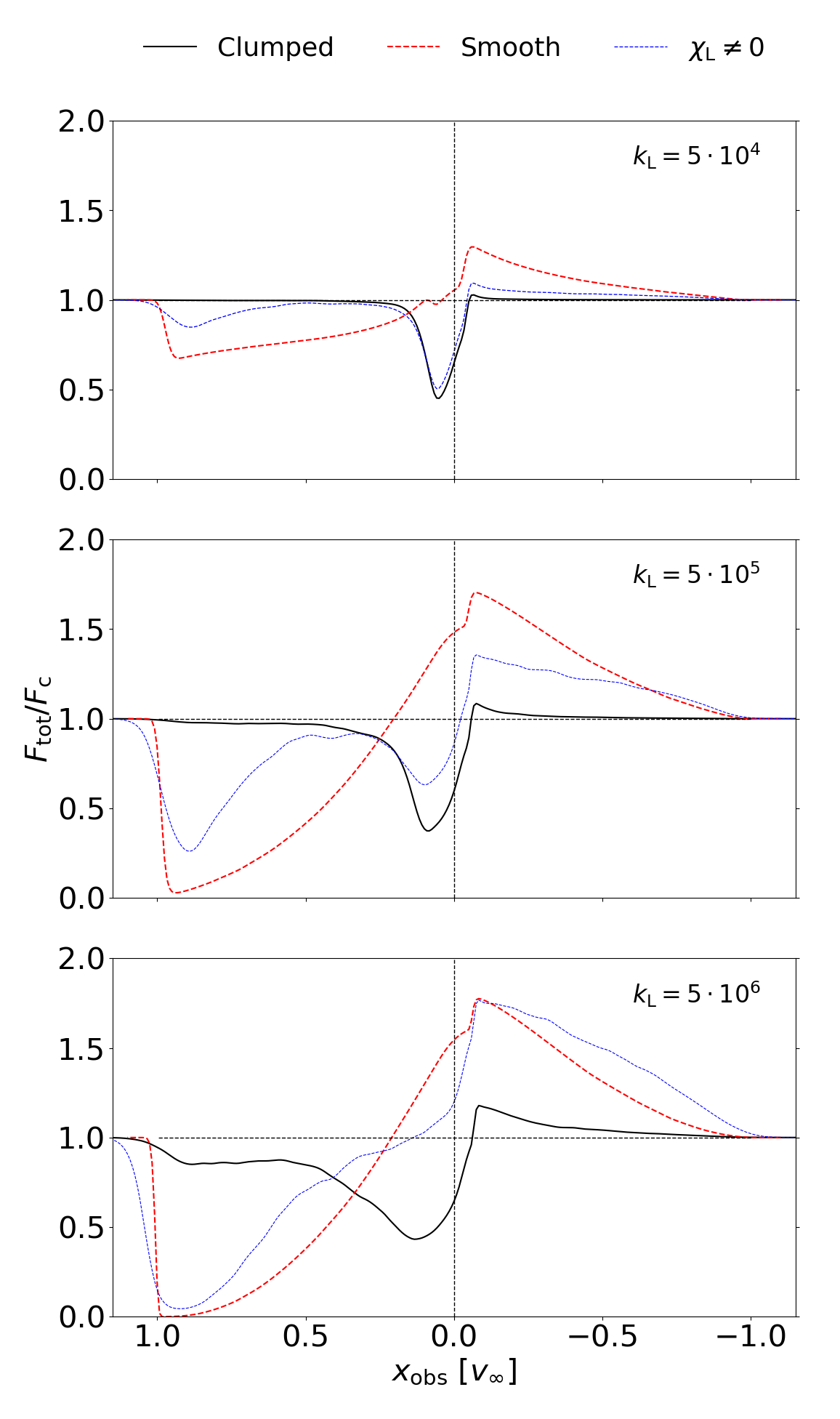}}  
%
%
\caption[]{Synthetic UV line profiles for the O-dwarf calculated from the patched LDI model (black solid lines) as compared to corresponding profiles obtained using a smooth $\beta-$velocity law wind model (red dashed lines). The top, middle, and bottom panels represent a `weak', `intermediate', and `strong' line with $\kline=[5\cdot10^4,5\cdot10^5,5\cdot10^6]$, respectively. The blue dotted lines represent the line profiles as obtained for the `clumped' model without accounting for the opacity cut-off at temperatures above $10^5$ K, thus showing only the velocity-porosity effect (see text).}
\label{fig:profiles}
\end{figure}
\begin{figure}[ht]
%
  \resizebox{\hsize}{!}{\includegraphics{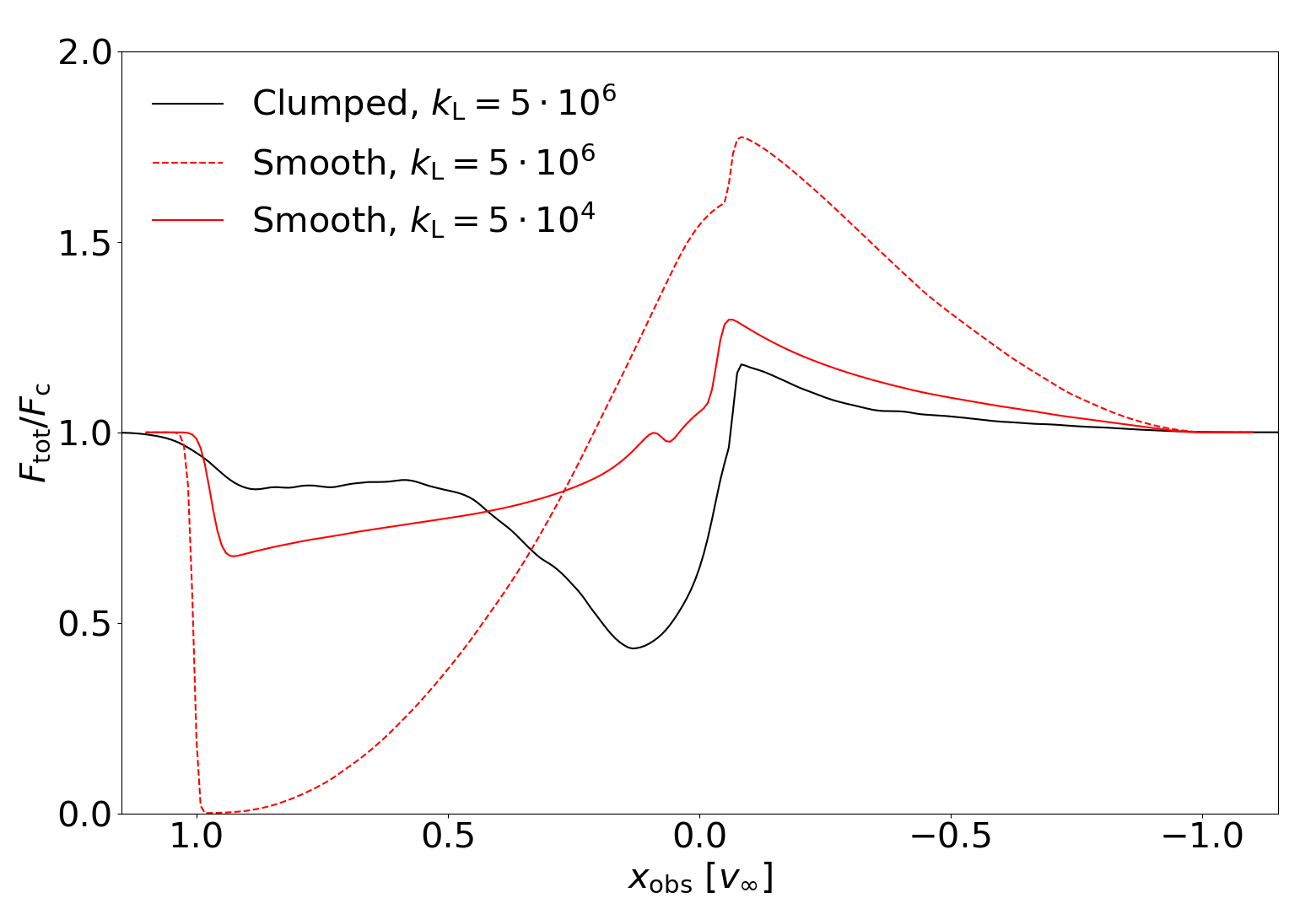}}
%
%
\caption[]{Synthetic UV line profile for the O-dwarf calculated from the patched LDI model using $\kline=5\cdot10^6$ (black solid line), as compared with corresponding line profiles obtained from the `smooth' wind model with $\kline=5\cdot10^6$ (red solid line) and $\kline=5\cdot10^4$ (red dashed line).}
\label{fig:profiles_smcl}
\end{figure}
%
%
%
%
%
%
Figure \ref{fig:profiles} shows synthetic UV line profiles computed directly from the low-density O-dwarf LDI model presented above using the short-characteristics method outlined in $\S$\ref{subsec:profiles}. The figure displays three different lines, characterised by line-strength parameters $\kline = [5\cdot10^4,5\cdot10^5,5\cdot10^6]$, which here are taken to represent weak, intermediate, and strong lines, respectively. In each panel, the profiles computed from the 3D snapshot-renderings of the simulations (black solid lines) are compared to corresponding profiles (red dashed lines) calculated from a smooth wind model. In order to mimic the line profiles as typically resulting from state-of-the-art 1D codes, the smooth wind model is determined by a $\beta-$velocity law and the equation of continuity, with the velocity stratification obtained from a best-fit model to the time-averaged velocity field of the LDI simulations, and the mass-loss rate set accordingly to the time-averaged mass-loss rate. Thus, the smooth wind models represent a monotonic velocity field without any shock-heated regions (and therefore no zones where the opacity has been cut are present).
As the average mass-loss rates of the `smooth' and `clumped' models are the same, any differences seen in the synthetic line profiles can be attributed to effects from the non-homogeneous wind structure and the hot regions. 

The figure clearly shows that, in particular for the lines of intermediate and strong strength, the synthetic profiles computed from the structured wind are significantly weaker than those emerging from the smooth model. This prominent line strength reduction stems from the fact that large regions in the structured wind are hot and thus have very low (or even zero) line opacities. The overall effect is then to reduce the average amount of line absorption, such that the lines appear weaker than they would if the wind was smooth and isothermal.

This `light-leakage' effect is similar to that produced by leakage of line photons from the resonance zones of a wind that is porous in velocity space \citep{oskinova07, Sundqvist11,Sundqvist2018}. Indeed, the two effects are here present simultaneously, because such `velocity porosity' is also a natural consequence of the typical structure emerging from the LDI \citep{owocki08, Sundqvist2010}. However, as demonstrated by the third set of profiles displayed in Fig. \ref{fig:profiles} (blue dotted lines), the effect of the hot gas is dominant for these low-density winds, particularly in the outer wind at high velocities. Namely, while these comparison profiles are computed from the identical `clumped' model as before, we have not applied any opacity cut-off within the hot wind regions for them. Therefore, the full wind now contributes to the line opacity (as if the simulation had been isothermal), making it possible to isolate the velocity-porosity effect. Furthermore, as seen in the figure, the reduction of opacity by the hot gas is of the same order (or even dominant in the outer wind) as the corresponding effect from velocity porosity that is calculated by omitting the temperature cut-off. Thus, accounting for the hot gas in such low-density winds plays a crucial role in determining empirical mass-loss rates.

We further note that the opacity reduction is most prominent for the line of intermediate strength. This is as expected, because for the weaker line large portions of the gas will in any case be relatively optically thin, thus reducing the overall net effect of the hot gas on the appearance of the line profile. However, for the intermediate line (which in the smooth wind is rather strong, but still unsaturated), the full opacity reduction within the hot gas affects the profile significantly. As such unsaturated lines are the ones typically used for deriving empirical mass-loss rates from observations, the importance of properly accounting for the hot gas in quantitative wind spectroscopy of main sequence OB-type stars becomes evident. 

Figure \ref{fig:profiles_smcl} demonstrates this explicitly, now comparing the same $\kline=5\cdot10^6$ line computed from the LDI simulation to two different profiles with $\kline=5\cdot10^4$ and $\kline=5\cdot10^6$ computed from the smooth outflows. The goal of this figure is not to provide a good fit but rather to qualitatively compare the absorption depths, which are an empirical measure for the mass-loss rate. Moreover, by looking at Fig. \ref{fig:profiles_smcl} it becomes clear that using a $\beta$-wind velocity model to fit observed UV lines is a poor method in the case of hot winds. In our comparison, the smooth $\kline=5\cdot10^4$ line is a much better match to the LDI model with a higher $\kline=5\cdot10^6$. For these UV scattering lines $\chil \propto \kline\dot{M}$ (see $\S$\ref{subsec:profiles}), and therefore this implies that, according to our simulations, a mass-loss rate derived from observational fitting of unsaturated line profiles using a model that disregards the direct opacity effects of the large volume of hot gas will be overestimated by a factor of $\approx10-100$. As further discussed in $\S$4 below, this may have important consequences for quantitative wind spectroscopy of main sequence OB-type stars, and in particular for the interpretation of the so-called weak-wind problem.

Finally, let us also note that the terminal velocity that would be measured from the weak and intermediate synthetic line profiles in Fig. \ref{fig:profiles} underestimates the actual simulated terminal velocity. Thus, our simulations here tentatively provide a natural, though only qualitative, explanation for the discrepancy between observed and theoretically predicted (from steady-state models) terminal velocities (see introduction, and also \citealt{Sundqvist19}).
  
\section{Discussion and conclusions}\label{Sec:conclusion}
We present the first radiation-hydrodynamic LDI simulation of a massive main sequence star in the `weak-wind' regime (see Sect. 1). The inherent issue of modelling radiative cooling in a supersonic radiating flow has been alleviated within the cooling curve implementation by relying on suggestions of previous work \citep{Feldmeier1995}. Subsequently, we applied a 3D short-characteristics radiative transfer code on the obtained model to produce synthetic UV line profiles and to analyse these in the context of the weak-wind problem.

The main result found from the hydrodynamic simulations is that radiation-driven weak winds are substantially shock heated by the LDI to high temperatures ($T>10^5$ K) and are unable to cool down efficiently. Specifically, for our template model we show that on average $69\pm10 \%$ of the wind volume can be considered hot. As such, our simulations provide basic support for some previous ideas of a primarily hot radiation-driven weak wind \citep[e.g.][]{Lucy2012}. 
Calculating synthetic UV line profiles directly from the shock-heated weak-wind model, we demonstrate that these are significantly weaker than those predicted by smooth wind models with the same mass-loss rate. Quantifying this, we find that the reduction in optical depth from the hot gas can be up to two orders of magnitude for an unsaturated UV wind line.

This result has important consequences for UV spectroscopy of weak-wind stars. Namely, in standard atmosphere and wind models ({\sc cmfgen}, \citealt{hilliermiller98, Hillier12}; {\sc PoWR}, \citealt{Sander17}, {\sc fastwind}, \citealt{Puls20}), the presence of hot gas is typically included only through its ionisation feedback from very small wind volumes, and not as a (here significant) modification of the bulk wind temperature structure. As such, the significant effect on UV lines found in this paper is currently never accounted for when inferring mass-loss rates by means of quantitative UV spectroscopy, which is the standard method for obtaining such empirical mass-loss rates in this weak-wind domain\footnote{because the optical H$\alpha$ line becomes insensitive. However, see discussion in \citealt{Puls08b} for a potential IR alternative and \citet{Najarro11} for a first application.}. 
In addition, one might perhaps also suspect that the abundance of hot gas and accompanying change in ionisation stages could make some lines stronger. However, to investigate this effect we would need a proper description of the detailed wind ionisation balance, which is beyond the scope of this paper (see also discussion below).

Thus, an important conclusion from our study is that radiation-driven weak winds might not be weaker than theoretically expected. Instead, we find that they have a complicated thermal structure where more than half of the wind volume can be significantly hotter than the stellar effective temperature, which means their actual mass-loss rates are significantly higher than what is derived empirically from the UV by means of standard non-local thermodynamic equilibrium (NLTE) model atmosphere codes. As just one specific example, \citet{Almeida19} derive $\dot{M}_{\mathrm{UV}} = 5 \times 10^{-9} \, \rm M_\odot/yr$ from an unclumped {\sc cmfgen} analysis of the Galactic weak winded O9 star HD\,156292 (see their Table 3), whereas the new mass-loss formula from \citet{Bjorklund20} (their eq. 20) provides a predicted $\dot{M}_{\mathrm{th}} = 3.5 \times 10^{-8} \ \rm M_\odot/yr$ for this star. This gives $\dot{M}_{\mathrm{UV}}/\dot{M}_{\mathrm{th}} \approx 0.14$, which agrees with the mass-loss underestimations typically expected for such standard UV line analyses (see $\S$\ref{Sec:results}). Therefore, the line-driven shock-heated model presented in this paper offers a tentative solution to the weak-wind problem.

Furthermore, we used the 3D short-characteristics radiative transfer code by \citet{Hennicker2020} to investigate the principal effect of the hot gas on the UV line opacity, using a simple high-temperature opacity cut-off and comparing to corresponding smooth models in order to examine the differential effects. However, in order to obtain fully quantitative results, it will be necessary to obtain a proper wind ionisation balance. Spectroscopic NLTE models currently in use should therefore be appropriately updated to account for the potentially large volumes of hot gas that might be present in weak winds. Only then can UV wind spectroscopy be used reliably in this regime.

Finally, some high-energy X-ray observations also seem to support the idea that the low-density winds of OB-type stars consist of large portions of hot gas \citep{Huenemoerder12, Doyle17}. We focused on the computation of the first LDI simulations of such low-density winds, and the subsequent analysis of UV spectral lines in the context of the weak-wind problem. In future work, it will be interesting to also calculate X-ray emissions from the weak-wind models presented here, and to quantitatively compare these to observational data.

\begin{acknowledgements}
We thank the referee for their constructive feedback on the paper. FAD, LH, and JOS acknowledge support from the Odysseus program of the Belgian Research Foundation Flanders (FWO) under grant G0H9218N. NDK acknowledges support from the KU Leuven C1 grant MAESTRO C16/17/007.
\end{acknowledgements}

%
%


\end{document}